\documentstyle[12pt]{article}
\newlength{\bredde}
\def\slash#1{\settowidth{\bredde}{$#1$}\ifmmode\,\raisebox{.15ex}{/}
\hspace*{-\bredde} #1\else$\,\raisebox{.15ex}{/}\hspace*{-\bredde} #1$\fi}
\newcommand{\beq}{\begin{equation}}
\newcommand{\eeq}{\end{equation}}

\def\gtwid{\raise.3ex\hbox{$>$\kern-.75em\lower1ex\hbox{$\sim$}}}
\def\ltwid{\raise.3ex\hbox{$<$\kern-.75em\lower1ex\hbox{$\sim$}}}
\begin{document}
\topmargin -0.8cm
\oddsidemargin -0.8cm
\evensidemargin -0.8cm
\headheight 0pt
\headsep 0pt
\topskip 9mm
\title{\Large{
The D=1 Matrix Model and the Renormalization Group}}

\vspace{1.2cm}

\author{
{\sc J. Alfaro}\thanks{Permanent address: Fac. de Fisica, Universidad
Catolica de Chile, Casilla 306, Santiago 22, Chile.} and
{\sc P.H. Damgaard}\\
CERN -- Geneva
}
\maketitle
\vfill
\begin{abstract} We compute the critical exponents of $d = 1$
string theory to leading order, using the renormalization group approach
recently suggested by Br\'{e}zin and Zinn-Justin.
\end{abstract}
\vfill
\vspace{11cm}
\begin{flushleft}
CERN--TH-6546/92 \\
June 1992
\end{flushleft}
\newpage


Recently, Br\'{e}zin and Zinn-Justin \cite{Brezin} have suggested a new
approach to the double scaling limit \cite{Brezin2} of matrix models
\cite{Ambjorn}. The idea is to take more seriously the notion that
the double scaling limit is similar to the approach to an
ordinary critical point, familiar from statistical physics and the
continuum limit of quantum field theories. Br\'{e}zin and Zinn-Justin
show that in principle one is
able to use the standard machinery of the Renormalization Group to analyze
very succinctly the behavior close to the scaling limit. Intuitively it
follows from the observation that a shift in $N$, the size of the matrix,
can be compensated by a shift in the coupling constant $g$. The size of
the matrix, $N$, then plays a r\^{o}le very similar to the cut-off
or inverse lattice spacing in
the conventional continuum limit of quantum field theories. Going to
the double scaling limit entails sending this `cut-off' $N$ to infinity
while keeping a certain scaling variable fixed.

The main motivation is not to rederive results that can be obtained by other
direct means for theories with $c < 1$. But if one is to make progress beyond
the apparent barrier at $c = 1$, it is important to have a general
calculational tool available that allows to treat
matrix models in different dimensionalities on an almost equal footing.
This, of course, refers in particular to models with
$c > 1$. It is widely believed that going beyond $c=1$ will, one way
or another, involve approximations \cite{Alvarez-Gaume}, rather than
exact solutions.

To see if a renormalization group approach to matrix models has the potential
to yield accurate results even in cases so far unexplored by conventional
means, it is important to know its possible shortcomings in known
surroundings. A test of this idea on models with known critical exponents
in the double scaling limit is therefore needed. This paper is one further
step in this direction.

One crucial ingredient of the renormalization group analysis by Br\'{e}zin
and Zinn-Justin is the existence of Callan-Symanzik equation of the form
\beq
\left[N \frac{\partial}{\partial N} - \beta(g)\frac{\partial}{\partial g}
+ \gamma(g) \right] Z(N,g) ~=~ r(g)
\eeq
which is satisfied by the string partition function $Z(N,g)$. The
$\beta$-function has a fixed point $g^*$ with $\beta'(g^*) = 0$ and
$\beta(g^*) > 0$. The singular part of the string partition function is then
of the scaling form
\beq
Z(N,g) ~=~ \Delta^{2 - \gamma_0} f(\Delta N^{2/\gamma_1})
\eeq
with $\Delta \equiv g^* - g$, and
\beq
\gamma_1 ~=~ \frac{2}{\beta'(g^*)}~,~~~~~\gamma_0 ~=~ 2 - \frac{\gamma(g^*)}
{\beta'(g^*)}.
\eeq
Consistency with the continuum theory
\cite{Knizhnik} requires
\beq
\gamma_1 = 2 - \gamma_0 = \frac{1}{12}\left[25 - c +
\sqrt{(1-c)(25-c)}\right]
\eeq
where $c$ is the central charge.

In the matrix model approach a renormalization group equation of the form (1)
is satisfied by the free energy (string partition function) if the matrix
model partition
functions $\zeta(N,g)$ and $\zeta(N-1,g')$ of the two
theories defined by $N\times N$ and $(N-1)\times (N-1)$ matrices, respectively,
are related by a recursion relation of the form
\beq
\zeta(N,g) ~=~ [\ell(g)]^{N^{2}} \zeta(N-1,g').
\eeq
with $\ell(g) = 1 + \frac{1}{N} r(g)$.

Br\'{e}zin and Zinn-Justin \cite{Brezin} have solved the resulting
renormalization group equations to leading order in perturbation theory
for the $c=0$ theory described by a one-matrix model. In that model, the
identity
\beq
\gamma_1 + \gamma_0 ~=~ 2
\eeq
is satisfied automatically, since the overall $N^{-2}$ factor in the
definition of the free energy
density implies $\gamma(g) = 2$. Considering that everything is
computed to lowest non-trivial order, the results of ref.
\cite{Brezin} are quite
encouraging: the exponents for the ordinary critical point are
$\gamma_1 = 2$ (and hence $\gamma_0 = 0$), which should be compared
with the exact $c=0$ values 5/2 and -1/2, respectively. At the $m$-th
multicritical point, the exponents become asymptotically exact.

The purpose of this short note is to extend the results of ref. \cite{Brezin}
to the $c=1$ matrix model. The scaling laws are known to be much more intricate
in this case, with logarithmic corrections to the pure power-law scaling
for $c < 1$ \cite{Kazakov,Brezin3}. Still, the power-law behavior is
correctly described by the formula (4) for $c=1$, which gives
$\gamma_1= 2-\gamma_0 =2$. Since we are only going to try to extract
the leading behavior, we can hope not to see the subtleties associated
with logarithmic corrections, and we therefore work on the
assumption that the
scaling form (2) holds in this case as well.

As in earlier work \cite{Kazakov,Brezin3}, we assume that the $c=1$ theory
can be represented by a one-dimensional matrix model,
\beq
\zeta(N,\mu,\lambda) ~=~ \int [dM] e^{-\int dt {\cal L}}
\eeq
with
\beq
{\cal L} ~=~ \frac{1}{2} Tr \dot{M}(t)^2 + \frac{1}{2}\mu^2 Tr M(t)^2
+ U(M)
\eeq
where $U(M)$ is the potential, and $M(t)$ is an $N\times N$ hermitian
matrix.
This theory is
usually solved by means of the central observation \cite{Brezin4} that
it can be related to a quantum mechanical theory of non-interacting
fermions in an external potential. Since
we are interested in observing the modifications due to explicitly
integrating out the degrees of freedom corresponding to going from
an $N\times N$ to an $(N-1)\times (N-1)$ matrix model, the original
formulation in terms of a one-dimensional matrix field theory is much
more convenient for our purposes.
Carlson \cite{Carlson} has studied the ordinary
large-$N$ limit of the same model using similar renormalization group
considerations. The idea is to
consider the expansion of the free energy, or the partition function
itself, in terms of
vacuum diagrams. The partial summation of degrees of freedom can then
be done directly in perturbation theory. To leading order in the
perturbative expansion no new operators appear in the new
effective action.

For definiteness, consider the case of a quartic potential:
$U(M) = (\lambda/N) TrM(t)^4$. Define the free energy $f(N,\mu,\lambda)$ by
\beq
e^{-N^2Tf(N,\mu,\lambda)} ~=~ \int [dM] \exp[-\int^T_0 dt {\cal{L}}]
\eeq
We now integrate out degrees of freedom corresponding to having one index
of the matrix $M(t)$ equal to $N$, using ordinary perturbation theory.
(Recall that the large-$N$ free energy is analytic around $\lambda = 0$.)
In detail:
\begin{eqnarray}
& & \int [dM] e^{-\int_0^T dt Tr[\frac{1}{2}\dot{M}^2 + \frac{\mu^2}{2}M^2]}
(1 - \frac{\lambda}{N} \int_0^T dt Tr M(t)^4 + \ldots) \nonumber \\
& \simeq & \zeta(N,\mu,0) \langle 1 - \frac{\lambda}{N} \int_0^T
Tr M(t)^4 \rangle.
\end{eqnarray}
To leading order in $1/N$, this is evaluated by just planar contractions,
{\em viz.:}
\beq
\langle Tr M^4 \rangle = \langle M_{ij}M_{ji} M_{kl}M_{lk} \rangle +
\langle M_{ij}M_{jk}M_{kl}M_{li} \rangle + {\cal{O}}\left(\frac{1}{N^2}\right)
\eeq
Computing, within perturbation theory, the right hand side for fixed indices
$i,j$ or $k$ equal to $N$ leads to\footnote{We differ in some details from
the analysis of ref. \cite{Carlson}.}
\beq
\langle Tr M^4 \rangle_N = \langle Tr M^4 \rangle_{N-1} + \frac{2}{\mu}
\langle Tr M^2 \rangle_{N-1} + {\cal{O}}(N^2)\cdot\langle 1 \rangle
\eeq
where the subscripts indicate whether the matrix indices run up to
$N$ or $N-1$. The term on the right hand side corresponding to a constant
operator is irrelevant for the computation of the critical
exponents, and we shall not be concerned with its actual value.

Next, we re-exponentiate the result (12) to get, up to $M$-independent terms:
\begin{eqnarray}
& & e^{-N^2Tf(N,\mu,\lambda)} ~=~  \nonumber \\
& & \int [dM]\exp\left\{-\int_0^T dt Tr[\frac{1}{2}\dot{M}^2
+ \frac{\mu^2}{2}M^2 + \frac{2\lambda}{N\mu}M^2 + \frac{\lambda}{N}M^4
+ \ldots ]\right\}
\end{eqnarray}
where the
matrix $M(t)$ on the right hand side is of size $(N-1)\times (N-1)$.
If we identify the renormalized couplings $\mu'$ and $\lambda'$
of the $(N-1)\times (N-1)$ theory as
\beq
\lambda' ~=~ (1 - \frac{1}{N})\lambda
\eeq
and
\beq
\mu' ~=~ \mu + \frac{2}{N}~\frac{\lambda}{\mu^2}
\eeq
then
\beq
N^2 f(N,\mu,\lambda) ~=~ (N-1)^2 f(N-1,\mu',\lambda') - N r(\mu,\lambda)
\eeq
where $r(\mu,\lambda)$ contains the constant terms from the action (13).

It follows that the free energy satisfies the renormalization group equation
\beq
\left[N\frac{\partial}{\partial N} - \beta_1(\lambda)\frac{\partial}
{\partial \lambda} - \beta_2(\mu)\frac{\partial}{\partial \mu} +
2 \right] f(N,\mu,\lambda) = r(g)
\eeq
with
\beq
\beta_1(\lambda) = - \lambda~,~~~~~~~\beta_2(\mu) = + \frac{2\lambda}{\mu^2}.
\eeq

For dimensional reasons, the free energy can always be written as either
$f(N,\mu,\lambda) = \mu Z(N,g)$ or, equivalently, $f(N,\mu,\lambda)
= \lambda^{1/3} \tilde{Z}(N,g)$, with a dimensionless coupling constant
\beq
g ~\equiv~ \frac{\lambda}{\mu^3}.
\eeq
The former representation is more convenient for our purposes. The
dimensionless string partition function $Z(N,g)$ defined as above then
satisfies a Callan-Symanzik equation of the form

\beq
\left[N \frac{\partial}{\partial N} - \beta(g)\frac{\partial}{\partial g}
+ \gamma(g) \right] Z(N,g) ~=~ r(g)
\eeq
with
\beq
\beta(g) ~=~ \left[\frac{\beta_1(\lambda)}{\mu^3} - \frac{3\beta_2(\mu)
\lambda}{\mu^4}\right] ~=~
- g - 6g^2
\eeq
and
\beq
\gamma(g) ~=~ 2 - \frac{\beta_2(\mu)}{\mu} ~=~ 2 - 2g.
\eeq

We thus find a non-trivial critical point at $g^* = - 1/6$, which should be
compared with the exact value $g_c = - \sqrt{2}/3\pi = - 0.15005\ldots$
\cite{Brezin4}. The
critical exponents follow from eq. (3), and turn out to be
\beq
\gamma_1 = 2~,~~~~~ \gamma_0 ~=~ 2g^* ~=~ - 1/3
\eeq
so that we get the exact result for $\gamma_1$ even in this first crude
approximation, but a value for $\gamma_0$
which differs from the exact exponent by a term of order $g^*$. Note that
this also implies that the relation (6) is no longer automatically
satisfied to this order in perturbation theory. In fact,
the identity (6) can only be recovered if the zero of
$\beta(g)$ is achieved by having both pieces of eq. (21) vanish
separately.
To this order, this occurs only at the trivial fixed point
at $g=0$. It follows from eq. (21) that it is precisely the violation
of the relation (6) that produces a non-trivial fixed point to this
order.\footnote{Clearly indicating the need to
go beyond this leading order approximation.}
This effect
can be better understood if one considers a slightly modified
renormalization procedure.

To this end, consider a scheme in which the mass term in the Lagrangian (8)
is normalized to one. We then have only one coupling constant, $\lambda$,
but of course need to consider also rescalings of the matrix fields in order
to maintain a fixed coefficient of the $Tr M^2$ operator. Computing in
perturbation theory as before, we find the leading order result (14) for
$\lambda$ again. But in order to fix the coefficient of the $Tr M^2$ to
be 1/2, we must perform a rescaling of the matrix,
\beq
M ~\to~ (1 + \frac{4\lambda}{N})^{-1/4}M  ~\equiv~ \rho(\lambda,N)^{-1/4}M
\eeq
and a simultaneous rescaling of time (in order that the $Tr \dot{M}^2$
operator remains with fixed coefficient as well),
\beq
t ~\to~ t\rho(\lambda,N)^{-1/2}.
\eeq
This means that the renormalized partition function is evaluated on
a time interval $[0,T\rho^{1/2}]$, while the original partition function
was evaluated on the interval $[0,T]$. The renormalization group equation
for the string partition function $Z(N,\lambda) =
(N^2T)^{-1}\ln\zeta(N,\lambda)$ then reads
\beq
\left[N\frac{\partial}{\partial N} - \beta(\lambda)\frac{\partial}
{\partial\lambda} + \gamma(\lambda)\right] Z(N,\lambda) = r_1(\lambda)
\eeq
where $r_1(\lambda)$, whose precise value is not of
interest here, arises from both the constant term in (12) and the jacobian from
the rescalings. Performing these rescalings, we find, to this order,
\beq
\beta(\lambda) ~=~ -\lambda - 6\lambda^2~,~~~~ \gamma(\lambda) ~=~ 2 - 2\lambda
\eeq
so that the $\beta$-function and the critical exponents indeed come out as
in the other scheme. We see now that
the shift in $\gamma(\lambda) \to 2 - 2\lambda$ instead of just
$\gamma(\lambda) = 2$ arises as an infrared effect: in this
formulation it is entirely due to the change in time interval $T \to
T\rho^{1/2}$. Thus, instead of the engineering scaling behavior $N^{-2}$,
we find to this order in perturbation theory an anomalous dimension.

Pushing this kind of renormalization group calculation of the $d=1$
theory to higher orders in perturbation theory will invariably introduce
non-localities in the effective action, rendering a standard Wilson-type
interpretation of the renormalized action more difficult. It is not
immediately obvious if these terms can be
approximated by local interactions in a manner reminiscent of mean field
theory \cite{Carlson}, without systematically altering the exponents obtained
in the double scaling limit.

The critical exponent $\gamma_1$ happens to agree with that obtained by
the same
method in the $c=0$ model, and it is worthwhile to see why this is so.
First, it follows trivially from (3) that this exponent is insensitive
to the precise value of the second term in the expansion of the
$\beta$-function\footnote{As long as it is non-vanishing, in order to have a
non-trivial critical point at this order in perturbation theory.}
(when truncated there),
since for a $\beta$-function of the form $\beta(g) = Ag + Bg^2$, we always
have $\beta'(g^*) = -A$ at the critical point. Second, consider the
expansion (12) in the case of a $d$-dimensional matrix model. The only
modifications are
\beq
\frac{2}{\mu} \langle Tr M^2 \rangle ~\to~ 4 \int \frac{d^dp}{(2\pi)^d}
\frac{1}{p^2 + \mu^2} \langle Tr M^2 \rangle
\eeq
and similar changes in the constant term (which we again can ignore here),
while the identification of $\lambda'$ in terms of $\lambda$ is unaffected.
Formally, this already implies that the first term in the expansion of the
$\beta$-function is the same in all dimensions, which in turn, by the
argument above, means that to the first non-trivial order in perturbation
theory one obtains the same exponent $\gamma_1$ (which happens to be exact at
$d=1$).\footnote{The argument is purely formal, because in higher
dimensions one must face also the usual field theory renormalization
associated with changes in the ultraviolet cut-off. This does not,
however, have any bearing on the cases $d=0$ and $d=1$.}
Evaluating the expressions above for $d=0$ we in fact reproduce the
shift in the $Tr M^2$-term from the corresponding calculation in the work
of \cite{Brezin}, which one can see as a check on the present computation.
What is effectively a large-$N$ vector theory calculation  -- the
integration over a row and a column of the $N\times N$ matrix theory --
can be substantially improved if one sums all bubble diagrams to leading
order in $1/N$ \cite{Brezin}. This changes the
${\cal{O}}(g^2)$ contribution to the $\beta$-function, but as we have seen
above, such a modification will not have any effect on the critical
exponent $\gamma_1$ to this order. This
holds in this $d=1$ theory as well.

\vspace{0.5cm}
\noindent
{\sc Acknowledgement:} We thank V. Kazakov for discussions.
The work of J.A. has been partially supported by
Fundaci\'on Andes no. C-11666/4, and an EC CERN Fellowship.


\newpage
\begin{thebibliography}{X}
\bibitem{Brezin}E. Br\'{e}zin and J. Zinn-Justin, {\em Renormalization
Group Approach to Matrix Models}, preprint LPTENS 92/19, SPhT/92-064.
\bibitem{Brezin2}E. Br\'{e}zin and V. Kazakov, Phys. Lett. {\bf B236} (1990)
144.
\newline M. Douglas and S. Shenker, Nucl. Phys. {\bf B335} (1990) 635. \newline
D.J. Gross and A. Migdal, Phys. Rev. Lett. {\bf 64} (1990) 127.
\bibitem{Ambjorn} J. Ambj{\o}rn, B. Durhuus and J. Fr\"{o}hlich, Nucl. Phys.
{\bf B257} (1985) 433. \newline
F. David, Nucl. Phys. {\bf B311} (1985) 45. \newline
V. Kazakov, Phys. Lett. {\bf 150B} (1985) 28.
\bibitem{Alvarez-Gaume}L. Alvarez-Gaum\'{e} and J.L.F. Barb\'{o}n,
CERN preprint CERN-TH-6464/92.
\bibitem{Knizhnik}V.G. Knizhnik, A.M. Polyakov and A.A. Zamolodchikov,
Mod. Phys. Lett. {\bf A3} (1988) 819. \newline
F. David, Mod. Phys. Lett. {\bf A3} (1988) 207. \newline
J. Ditsler and H. Kawai, Nucl. Phys. {\bf B231} (1989) 509.
\bibitem{Kazakov}V. Kazakov and A. Migdal, Nucl. Phys. {\bf B311} (1989) 171.
\bibitem{Brezin3}E. Br\'{e}zin, V. Kazakov and A.A. Zamolodchikov,
Nucl. Phys. {\bf B338} (1990) 673. \newline
P. Ginsparg and J. Zinn-Justin, Phys. Lett. {\bf B240} (1990) 333. \newline
D.J. Gross and N. Miljkovic, Phys. Lett. {\bf B238} (1990) 217. \newline
G. Parisi, Phys. Lett. {\bf B238} (1990) 209.
\bibitem{Brezin4}E. Br\'{e}zin, C. Itzykson, G. Parisi and J.B. Zuber,
Comm. Math. Phys. {\bf 59} (1978) 35.
\bibitem{Carlson}J.W. Carlson, Nucl. Phys. {\bf B248} (1984) 536.

\end{thebibliography}
\end{document}